# Direct observation of quadruple spin-texture locking in a 2D *d*-wave altermagnet

Dan Mu[1, 2, *], Bei Jiang[1, 3, *], Qingchen Duan[1, *], Zulin Xu[1], Xingkai Cheng[4], Yusen Xiao[1], Xinru Han[1, 5], Xinyu Liang[1], Zhaokun Luo[1], Ryan L. Kong[1], Qiheng Wang[1], Junwei Liu[4], Jianxin Zhong[2, †], Ruidan Zhong[1, 6, †], Qiangqiang Gu[1, †], Baiqing Lv[1, 3, 7,†], Hong Ding[1, 8, 9]

[1] *Tsung-Dao Lee Institute & School of Physics and Astronomy, Shanghai Jiao Tong University, Shanghai 201210 China*

[2] *Institute for Quantum Science and Technology, Shanghai University, Shanghai 200444, China*

[3] *Shanghai Research Center for Quantum Sciences, Shanghai 201315, China*

[4] *Department of Physics, The Hong Kong University of Science and Technology, Hong Kong, China*

[5] *Department of Physics, Southern University of Science and Technology, Shenzhen 518055, China*

[6] *State Key Laboratory of Micro-nano Engineering Science, Shanghai Jiao Tong University, Shanghai 200240, China*

[7] *Zhangjiang Institute for Advanced Study, Shanghai Jiao Tong University, Shanghai 200240, China*

[8] *Hefei National Laboratory, Hefei 230088, China*

[9] *New Cornerstone Science Laboratory, Tsung-Dao Lee Institute, Shanghai Jiao Tong University, Shanghai 200240, China*

[*]These authors contributed to this work.

[†]Corresponding authors: jxzhong@shu.edu.cn; rzhong@sjtu.edu.cn; qiangqianggu2016@gmail.com; baiqing@sjtu.edu.cn

**Altermagnets combine vanishing net magnetization with nonrelativistic, momentum-dependent spin splitting**[1-5]**, offering a new paradigm for spintronics**[6,7]**. Spin-crystal symmetry coupling, namely spin-lattice locking, is the defining mechanism of altermagnetism, enforcing opposite spin sublattices in real space and spin-momentum-locked electronic structure in reciprocal space**[1-5]**. Direct atomic-scale visualization of spin-lattice locking therefore constitutes a decisive benchmark of the altermagnetic state, yet such evidence has remained elusive despite extensive efforts. Here we show that the electronic states in $RbV_2Se_2O$ exhibit a *d*-wave-like spin texture at the sublattice level, providing the first atomic-scale evidence of spin-lattice locking with a predominantly *c*-axis spin orientation. By employing an *in-situ*, field-switchable spin-polarized Cr tip, we realize spin-contrast mapping of quasiparticle interference at identical energies, overcoming a long-standing experimental barrier in altermagnets. The resulting interference patterns exhibit pronounced spin-dependent modulations, establishing spin-scattering locking and spin-momentum locking as the real and reciprocal space manifestations. Unexpectedly, we uncover that the spin-selective scattering response is organized by a long-period stripe modulation, giving rise to a previously unidentified form of spin-texture locking, spin-stripe locking. We attribute this behavior to the emergence of a spin-density-wave moiré pattern. Together, these results establish a unified picture of quadruple spin-texture locking phenomena in a *d*-wave altermagnet, and position altermagnets as a versatile platform for exploring many-body interactions among intertwined degrees of freedom, including spin, lattice, momentum, moiré potential and valley.**

Altermagnetism has recently emerged as a compelling unconventional magnetic phase[8-11] that bridges the regimes of ferromagnets and antiferromagnets, uniting the robustness of compensated magnetic order with the functionality of spin-polarized states[1-5]. This distinctive combination arises from an intricate interplay between magnetic order and crystal symmetry[12-15]: despite exhibiting collinear antiparallel arrangements reminiscent of conventional antiferromagnets, altermagnets host anisotropic spin densities on opposing sublattices that are related not by translation or inversion, but by mirror, rotational, glide, or screw symmetries[1-5]. Crucially, these real-space symmetries generate a momentum-dependent spin splitting that alternates across the Brillouin zone, reaching electron-volt energy scales even in the absence of spin-orbit coupling[16-24]. This real and reciprocal space alternating spin polarization renders altermagnets promising platforms not only for exploring fundamental quantum phenomena, including spin-triplet superconductivity[25-27], Majorana zero modes without net magnetization[28-30], and orientation-selective $\pi$-phase Josephson junctions[31-35], but also for developing spintronic devices with negligible stray fields and ultrafast dynamics[4,5,36-42].

Spin-crystal symmetry interaction, hereafter spin-lattice locking, underpins altermagnetism[1-3,12-15], engendering alternating spin polarizations in both real and reciprocal space[16-24,43-46] (Fig. 1a). In real space, a hallmark manifestation is spin-selective impurity scattering, wherein defect-induced standing-wave patterns are intimately tied to the real-space arrangement of anisotropic local spin densities, constituting spin-scattering locking. Correspondingly, in momentum space, it manifests as spin-momentum locking, with spin polarization locked to crystal momentum on constant-energy contours via $d$-wave, $g$-wave, or other even-parity symmetries[1-5]. Experimentally, spin-polarized scanning tunnelling microscopy[47] (SP-STM) represents the most direct approach for simultaneously accessing spin-lattice locking and its dual manifestations. Its capacity for atomic-scale spin-resolved imaging enables direct visualization of both spin-lattice and spin-scattering locking. Meanwhile, spin-momentum locking can be inferred from quasiparticle interference[48] (QPI) patterns arising from spin-selective scattering between states within the same spin channel, namely $E_\uparrow(\boldsymbol{k})$ and $E_\uparrow(\boldsymbol{k+q})$, or $E_\downarrow(\boldsymbol{k})$ and $E_\downarrow(\boldsymbol{k+q})$, where $E$ represents a constant-energy contour and ↑, ↓ label the spin channels.

However, resolving spin-texture locking with SP-STM—particularly spin-lattice locking—remains inherently challenging. Three stringent conditions must be satisfied. First, the measurement demands a reversibly switchable, spin-selective probe that remains robust under complex surface conditions. Second, access to the locking in both real and momentum space requires a hard-gap material that opens a clean low-energy window, so that impurity-bound states can be distinguished from the itinerant spin continuum. Third, the spin splitting must be sufficiently large to generate well-separated spin-preserved scattering wavevectors for the two spin channels. Consequently, direct visualization of spin-lattice locking, the microscopic basis of altermagnetism, and its dual-space manifestations, i.e., constant-energy spin-scattering locking and spin-momentum locking, has remained elusive.

In this work, leveraging SP-STM with an *in-situ* field-switchable Cr tip, we provide the first experimental evidence of quadruple spin-texture locking in the two-dimensional (2D) *d*-wave altermagnet RbV$_2$Se$_2$O. At constant energy, impurity-induced orthogonal standing waves with opposite spin polarization directly establish spin-scattering locking in real space. At the atomic scale, defect-induced electronic states at inequivalent V sites exhibit *d*-wave spin polarizations tied to the underlying lattice registry, providing the first experimental evidence for the long-sought spin-lattice locking in altermagnets. In momentum space, QPI measurements further reveal a characteristic *d*-wave spin polarization of the constant-energy contour near $E_F$, demonstrating spin-momentum locking. Surprisingly, we uncover a long-period stripe modulation accompanied by spin-stripe locking, in which defect-induced standing waves on neighboring stripes are locked to opposite spin polarizations. These findings provide a direct and exclusive demonstration of *d*-wave altermagnetism in RbV$_2$Se$_2$O, and highlight *d*-wave altermagnets as a fertile platform for investigating how spin textures intertwine with and lock to multiple degrees of freedom.

**RbV$_2$Se$_2$O as a canonical system for visualizing spin-texture locking**

Figure 1a schematically illustrates three intertwined forms of spin-texture locking in a 2D *d*-wave altermagnet: impurity-driven spin-scattering locking in real space, symmetry-enforced spin-lattice locking of the local spin density, and spin-momentum locking in *k* space associated with spin-split bands. RbV$_2$Se$_2$O provides a canonical system for visualizing this emergent spin-texture locking. The compound crystallizes in a layered tetragonal structure (upper-left inset of Fig. 1b), with Se atoms residing above and below the center of each V$_2$O square and Rb atoms intercalated between adjacent V$_2$Se$_2$O layers. Central to the magnetism is the V$_2$O plane, where alternating spin-up and spin-down V sublattices yield zero net magnetization, forming the basis of the altermagnetic state with a direction-dependent spin texture in both real space and momentum space (Extended Data Fig. 1). Cleavage exposes a uniform Rb terminated surface over a large field-of-view (FOV) (Fig. 1b). At atomic resolution, the bottom-right inset of Fig. 1b resolves a well-ordered $\sqrt{2} \times \sqrt{2}$ reconstruction of Rb layer, coexisting with a pronounced unidirectional stripe modulation. Figure 1c shows the spatial evolution of d$I$/d$V$($r$, $V$) measured along the trajectory indicated in Fig. 1b. The spectra exhibit a hard gap of approximately 50 meV, characteristic of the spin-density-wave (SDW) state[20,44-46,49,50]. The resulting suppression of itinerant spectral weight near the Fermi level opens a clean low-energy window, in which the intrinsic impurity-induced bound-state scattering can be resolved with minimal interference from itinerant spin channels.

**Spin-scattering locking**

Direct verification of spin-texture locking requires separating the two spin channels with a spin-sensitive probe. We therefore performed SP-STM measurements using Cr tips whose apex spin polarization can be switched and stabilized by an external magnetic field (Figs. 2 a, c). The spin-resolved conductance map $g_\uparrow$($r$, $V$ = -10 mV) (Fig. 2b) and $g_\downarrow$($r$, $V$ = -10 mV) (Fig. 2d) were measured using a $+z$-polarized Cr tip at $B$ = 2 T and a $-z$-polarized Cr tip at $B$ = -2 T, respectively. Elongated QPI modulations develop along the two orthogonal crystallographic $a$- and $b$-axes for the in-gap bound state, and recover the fourfold symmetric scattering features out of the gap (Extended Data Fig. 2). The joint histogram of $g_\uparrow$($r$, $V$ = -10 mV) versus $g_\downarrow$($r$, $V$ = -10 mV) in Fig. 2e, demonstrates that enhanced local density of states (LDOS) modulations are tied to stronger local spin polarization. More explicitly, the spin-resolved scattering is visualized through the spin-contrast map:

$$g_{\uparrow\text{-}\downarrow}(\boldsymbol{r}, V) = g_\uparrow(\boldsymbol{r}, V) - g_\downarrow(\boldsymbol{r}, V) \qquad (1)$$

The high-precision registration procedure (Methods and Extended Data Fig. 3) yields a correlation coefficient of 0.94, confirming that the two topographs are nearly identical and thus allowing a reliable extraction of the spin-contrast map. At constant voltage $V$ = -10 mV, $g_{\uparrow\text{-}\downarrow}$($r$, $V$ = -10 mV) displays red and blue contrast corresponding to spin-up and spin-down polarization, respectively. Thus, the one-to-one correspondence between scattering orientation and spin polarization directly reveals spin-scattering locking in real space.

**Spin-lattice locking**

The defining evidence for symmetry-enforced spin-lattice locking lies at the sublattice level. We therefore next performed spin-sensitive atomic-resolution imaging to uncover its microscopic manifestation. In the current channel, the signal reflects the energy-integrated LDOS,

$$I(\mathbf{r}, V) \propto \int_0^{eV} LDOS(\mathbf{r}, \varepsilon) d\varepsilon \tag{2}$$

The measured $I(\mathbf{r}, V = -10 \text{ mV})$ (Fig. 3a) exhibits the same elongated QPI modulations observed in Figs. 2b, d, while $I(\mathbf{r}, V = 20 \text{ mV})$ further reveals a pronounced four-lobed LDOS perturbation centered on an isolated defect (Fig. 3b). The high-resolution map resolves the Se lattice of a lattice constant $\sim 4$ Å, and identifies the defect on a Se site (inset of Fig. 3b), as illustrated in the local structure of the $V_2Se_2O$ layer in Fig. 3c. Most crucially, the atomic-scale spin-contrast current map, $I_{\uparrow-\downarrow}(\mathbf{r}, V = -10 \text{ mV}) = I_\uparrow(\mathbf{r}, V = -10 \text{ mV}) - I_\downarrow(\mathbf{r}, V = -10 \text{ mV})$ shows that spin-up and spin-down contrast reside predominantly on the two inequivalent V sublattices, $V_x$ and $V_y$, respectively (Fig. 3d). This sublattice-resolved spin polarization provides compelling evidence for intrinsic spin-lattice locking in $RbV_2Se_2O$, one of the central findings of this work. It is also worth highlighting that, given the out-of-plane orientation of the tip polarization, the observed spin contrast demonstrates that the magnetic moments at the V sites are predominantly aligned along the crystallographic $c$-axis. The above results provide the first atomic-scale evidence for both spin-lattice locking and the spin orientation of the V sublattices.

**Spin-momentum locking**

The real-space locking identified above naturally motivates an examination of its momentum-space band structure. In STM, this information is not accessed directly in $k$ space, but through Fourier transform (FT) QPI in $q$ space, where the scattering wavevector $\mathbf{q}$ connects the initial and final electronic state ($\mathbf{q} = \mathbf{k}' - \mathbf{k}$). Figure 4a shows $g_\uparrow(\mathbf{q}, V = -10 \text{ mV})$, the Fourier transform of $g_\uparrow(\mathbf{r}, V = -10 \text{ mV})$ in Fig. 2b. The QPI pattern displays a twofold anisotropy, with dominant scattering along $q_y$ rather than $q_x$. By contrast, the spin-averaged map (Extended Data Fig. 4) shows comparable scattering intensity along $q_x$ and $q_y$, demonstrating that the observed anisotropy is not caused by the tip artifact, but instead originates from spin-selective tunnelling. In Fig. 4b, the counterpart $g_\downarrow(q, V = -10\text{ mV})$, obtained from Fig. 2d, shows the complementary anisotropy with enhanced scattering along $q_x$. To quantify the difference between the two spin scattering channels, we define the spin-contrast QPI pattern $g_{\uparrow-\downarrow}(q, V) = g_\uparrow(q, V) - g_\downarrow(q, V)$. As shown in Fig. 4c, $g_{\uparrow-\downarrow}(q, V = -10\text{ mV})$ resolves both the small-$q$ ($q_{0x}$, $q_{0y}$) and the large-$q$ ($q_x$, $q_y$) scattering vectors, consistent with the spin-selection map analysis (Extended Data Fig. 5). Furthermore, the spin-dependent scattering is captured by the calculated spin-resolved QPI intensity in Fig. 4d using the $k_\parallel$-filtering procedure (Methods and Extended Data Fig. 6), which accounts for the momentum selectivity of STM tunnelling because states with larger in-plane momentum decay more rapidly into the vacuum barrier and contribute less to the measured QPI intensity. The striking agreement between experiment and calculation (Figs. 4c, d) confirms that the spin-resolved scattering channels are a direct consequence of spin-momentum locking in $RbV_2Se_2O$.

**Spin-stripe locking**

Finally, we demonstrate that these three distinct manifestations of spin-texture locking established in Figs. 2-4 converge into a unified picture, in which the spin texture is locked to the unidirectional stripe modulation. Figure 5a shows the topographic image $T(r)$, revealing the stripe pattern together with several typical scattering centers. The measured $g(r, V = 100\text{ mV})$ in the identical FOV as in Fig. 5a exhibits a nearly uniform unidirectional charge-stripe modulation. At lower energy around the gap edge, however, this apparently uniform stripe state breaks into two alternating inequivalent stripes, labelled A and B in Fig. 5c. These neighboring stripes are distinguished not simply by their contrast, but by their association with orthogonal scattering directions, indicating that the stripe modulation already encodes a hidden spin-dependent structure. To make this connection explicit, we use a 2D lock-in analysis to extract the local amplitude of the spin polarization,

$$A_q(r) = \int d\mathbf{R} A(\mathbf{R}) e^{i\mathbf{q}\cdot\mathbf{R}} e^{-\frac{(r-R)^2}{2\sigma^2}} \tag{3}$$

where A($\mathbf{R}$) denotes a real-space image, $\mathbf{q}$ is the wavevector of interest, and σ sets the averaging length scale in real space, or equivalently the momentum-space cutoff $\sigma_q = 1/\sigma$ (Methods). Figure 5d displays the impurity-induced local spin polarization, extracted from the lock-in amplitude of the spin-selection maps with $\sigma_q = 1\ nm^{-1}$ (Methods and Extended Data Figs. 5a, c), overlaid on the stripe modulations. Notably, adjacent A and B stripes carry opposite spin polarization, showing that the local spin texture is intrinsically locked to the unidirectional stripe modulation. A possible interpretation of these observations is illustrated schematically in Fig. 5e as an SDW moiré emerging from the doubled periodicity (2λ) of the unidirectional charge-stripe modulation. Under the altermagnetic background, the spin degree of freedom is largely pinned to the local quantization axis. The impurity potential may therefore be reduced to the diagonal form in the spin $\{|\uparrow\rangle, |\downarrow\rangle\}$ basis,

$$\hat{V} = U\hat{I} + M_z \sigma_z = \begin{pmatrix} U + M_z & 0 \\ 0 & U - M_z \end{pmatrix} \tag{4}$$

where $U$ denotes the potential scattering and $M_z$ the spin-dependent magnetic scattering. Accordingly, the nonzero impurity scattering matrix elements are

$$\langle \mathbf{k'}, \uparrow |\hat{V}| \mathbf{k}, \uparrow \rangle = U + M_z,\ \langle \mathbf{k'}, \downarrow |\hat{V}| \mathbf{k}, \downarrow \rangle = U - M_z \tag{5}$$

In this scenario, each pair of neighboring stripes constitutes a single magnetic unit, with adjacent stripes carrying opposite local magnetization. Such an alternating magnetic environment naturally modifies the spin-dependent scattering potential from stripe to stripe, so that the effective magnetic term is $M_z$ on one stripe and $-M_z$ on the next. As a result, the spin-conserving scattering amplitudes become strongly unequal if $U$ and $M_z$ are comparable (Methods): on one stripe, $|U + M_z| \gg |U - M_z|$ whereas on the neighboring stripe the relation is reversed, $|U + M_z| \ll |U - M_z|$. This alternating imbalance in the two spin-preserving scattering channels provides a natural explanation for the observed spin-stripe locking in Figs. 5c, d.

**Discussion and outlook**

Our results reveal a coherent hierarchy of spin-texture locking in RbV$_2$Se$_2$O. From impurity-induced scattering in real space, to sublattice-resolved spin structure, to anisotropic spin-dependent band dispersion in momentum space, the spin degree of freedom is consistently constrained across all length scales. This unified picture establishes a direct visualization of symmetry-protected spin textures in an altermagnet and provides a new paradigm for exploring spin-resolved electronic structure in correlated quantum materials.

The emergence of the long-period stripe order in RbV$_2$Se$_2$O is likely rooted in the reconstructed electronic structure imposed by the $\sqrt{2} \times \sqrt{2}$ SDW. Specifically, the SDW folds the Fermi pockets at the X and Y[18], after which the folded bands hybridize and open gaps. This reconstruction reshapes the Fermi surface and generates new nesting conditions with a much smaller wave vector. Such a cascade of electronic instabilities favors a secondary transition into a stripe order with a much longer period. In this sense, the stripe order and the primary $\sqrt{2} \times \sqrt{2}$ SDW can be viewed as forming an SDW morié superlattice, in which an emergent modulation develops on top of the original SDW background (Fig. 5e). Importantly, neighboring stripes carry opposite local net magnetizations, which break the equivalence between the two spin channels for impurity-induced bound-state scattering. As a result, one spin channel is preferentially selected on a given stripe, giving rise to both the pronounced twofold real-space spin-scattering locking and the spin-stripe locking observed in our SP-STM measurements.

By contrast, electronic states at Se-site defect exhibiting nearly fourfold-symmetric patterns are unlikely to promote such cascaded ordering and are more plausibly associated with Se vacancies. These defects locally perturb both strain and stoichiometry, thereby suppressing the SDW order in their vicinity. Nevertheless, such Se vacancies play a distinct and fortunate role in our measurements. Because they are located at the center of inequivalent V sites, the defect-induced states inherit the spin polarization tied to the underlying lattice registry, allowing the spin-lattice locking to be directly imaged at the atomic scale.

Looking forward, our observations of quadruple spin-texture locking in RbV$_2$Se$_2$O establish a unified picture of spin-symmetry formalism at atomic-scale and enable the resolution of 2D symmetry-protected spin configurations across both real and reciprocal space, providing a broadly applicable experimental framework that extends beyond altermagnets to spin-compensated magnets with nontrivial spin space groups. The 2D $d$-wave altermagnetic state identified in RbV$_2$Se$_2$O is particularly attractive because its non-reversing out-of-plane spin polarization is naturally suited to device architectures and the generation of spin currents, making it a promising platform for 2D spintronic applications. More broadly, our work highlights altermagnets as a fertile platform for exploring many-body interactions emerging from the interplay of multiple intertwined degrees of freedom, including spin, lattice, momentum, moiré potential and valley. All these directions await further investigation and are expected to stimulate broad interest in both fundamental studies and device-oriented applications of novel magnetic materials.

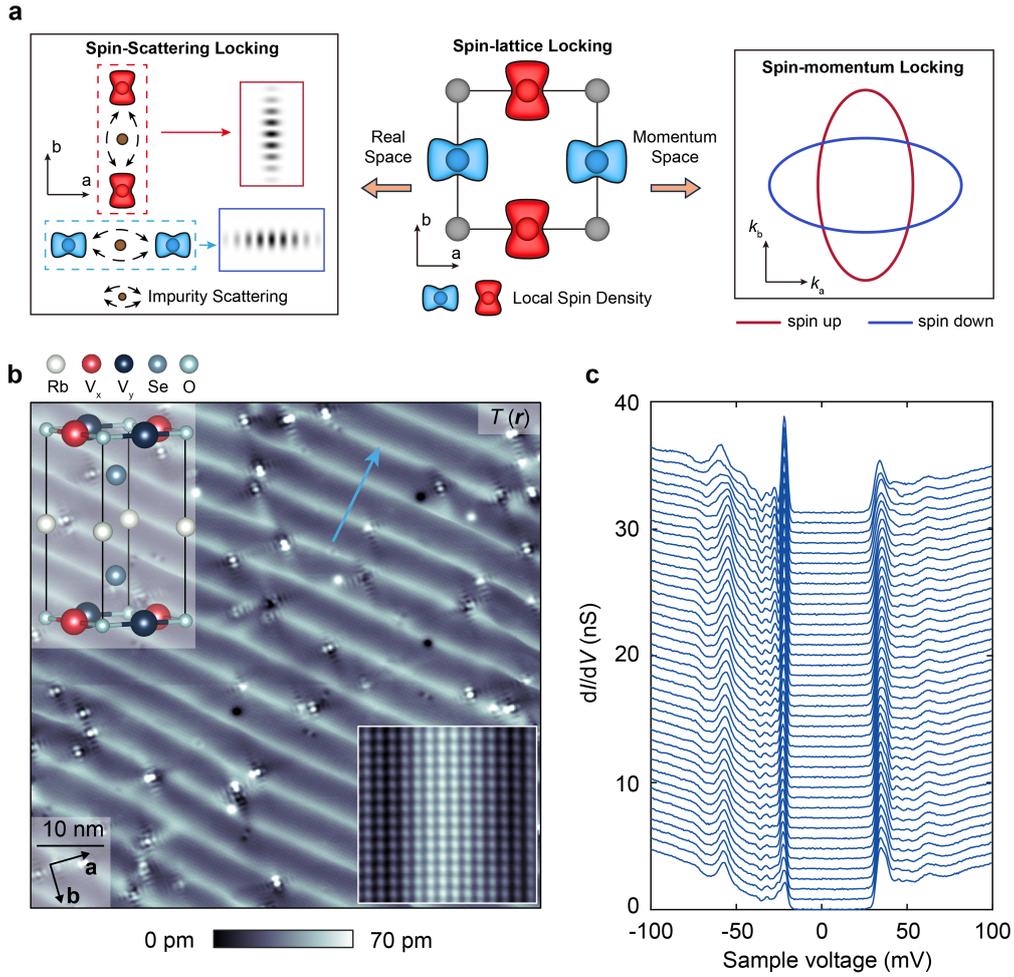

**Fig. 1 Emergent spin-texture locking in a 2D *d*-wave altermagnet: RbV$_2$Se$_2$O as a canonical system. a,** Schematic of the three intertwined manifestations of spin-texture locking in RbV$_2$Se$_2$O: impurity-driven spin-scattering locking in real space, symmetry-enforced spin-lattice locking of the local spin density, and spin-momentum locking in momentum space arising from spin-split bands. **b,** Typical topographic image *T*(***r***) of the cleaved RbV$_2$Se$_2$O surface, revealing a long-range unidirectional stripe modulation ($I_s$ = 0.3 nA, $V_s$ = 100 mV). Top-left inset: RbV$_2$Se$_2$O crystal structure; Bottom-right inset: atomically resolved topograph ($I_s$ = 0.2 nA, $V_s$ = 100 mV) exhibiting a well-ordered $\sqrt{2} \times \sqrt{2}$ reconstruction of Rb layer (Extended Data Fig. 1). **c,** Spatial evolution of the measured d*I*/d*V*(***r***, *V*) along the light-blue trajectory indicated in **b** ($I_s$ = 0.2 nA, $V_s$ = 100 mV). A hard gap of approximately 50 meV is observed in the SDW state, enabling direct visualization of impurity-induced bound-state scattering with minimal admixture from itinerant spin-up and spin-down states.

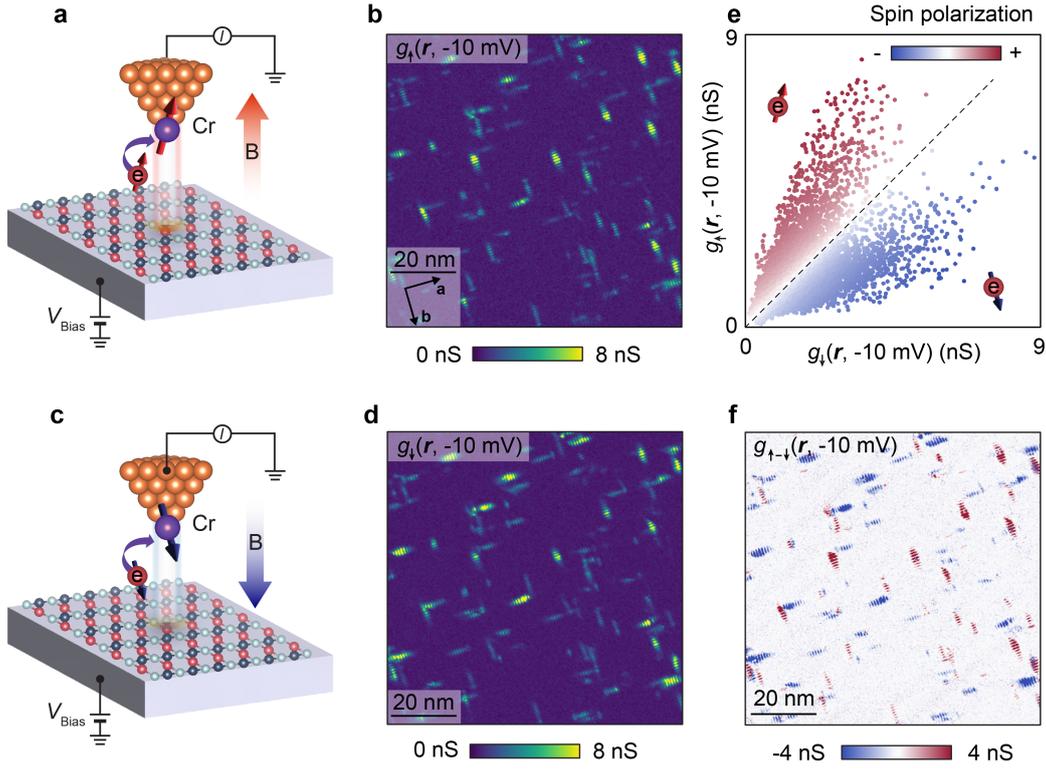

**Fig. 2 Real space imaging of spin-scattering locking in RbV$_2$Se$_2$O. a,** Schematic illustration of spin-selective tunneling under an external magnetic field applied along +z, which stabilizes a Cr tip with apex spin polarization along +z. **b,** Spin-resolved conductance map $g_\uparrow(r, V = -10$ mV$)$ acquired with the +z-polarized Cr tip at $B = 2$ T ($I_s = 0.3$ nA, $V_s = 100$ mV). Elongated QPI develops along the two orthogonal crystallographic $a$- and $b$-axes. **c,** Schematic illustration of spin-selective tunneling under an external magnetic field applied along −z, which stabilizes a Cr tip with apex spin polarization along −z. **d,** Spin-resolved conductance map $g_\downarrow(r, V = -10$ mV$)$ acquired with the −z-polarized Cr tip at $B = -2$ T ($I_s = 0.3$ nA, $V_s = 100$ mV), in the same FOV as in Fig. 2b. **e,** Joint histogram of $g_\downarrow(r, V = -10$ mV$)$ versus $g_\uparrow(r, V = -10$ mV$)$, characterizing the local spin polarization at each position $r$. Regions of enhanced LDOS modulation are found to exhibit larger spin polarization. The dashed line denotes the boundary between spin-up and spin-down polarization. **f,** Spin-contrast conductance map $g_{\uparrow-\downarrow}(r, V = -10$ mV$) = g_\uparrow(r, V = -10$ mV$) - g_\downarrow(r, V = -10$ mV$)$. The red (blue) contrast corresponds to spin-up (spin-down) polarization. The one-to-one correspondence between scattering orientation and spin polarization reveals spin-scattering locking in real space.

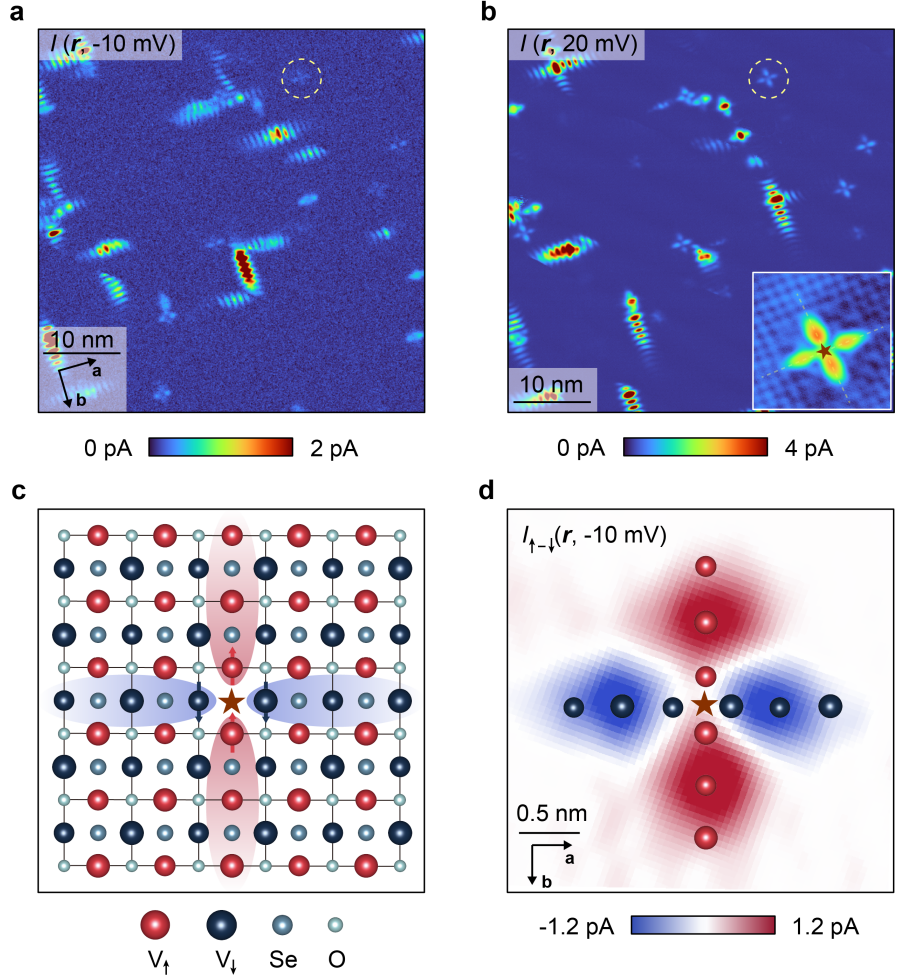

**Fig. 3 Atomic-resolution imaging of spin-lattice locking. a,** Measured current map $I(r, V = -10$ mV). Two orthogonal sets of elongated QPI modulations are observed, consistent with those in $g(r, V)$ in Figs. 2b, d. **b,** Measured current map $I(r, V = 20$ mV), revealing the same QPI modulations, along with the distinct fourfold-symmetric LDOS perturbations. Inset: High-resolution $I(r, V = 20$ mV) resolving the Se lattice (~ 4 Å), indicating that tunneling occurs through the underlying Se layer. A Se-site defect (marked by a star) is identified at the center of the LDOS perturbation. **c,** Lattice structure of the $V_2Se_2O$ layer viewed along (001). A fourfold-symmetric lobe pattern is centered at a Se-site defect. **d,** Zoomed-in spin-contrast current map $I_{\uparrow-\downarrow}(r, V = -10$ mV) $= I_{\uparrow}(r, V = -10$ mV) $- I_{\downarrow}(r, V = -10$ mV) at atomic scale, overlaid with $V_x$ and $V_y$ atomic positions. Spin-up (spin-down) contrast resides on the $V_x$ ($V_y$) sites, demonstrating spin-lattice locking to the two V sublattices. All measurements in this figure were performed with a setpoint of $I_s = 0.5$ nA and $V_s = 100$ mV.

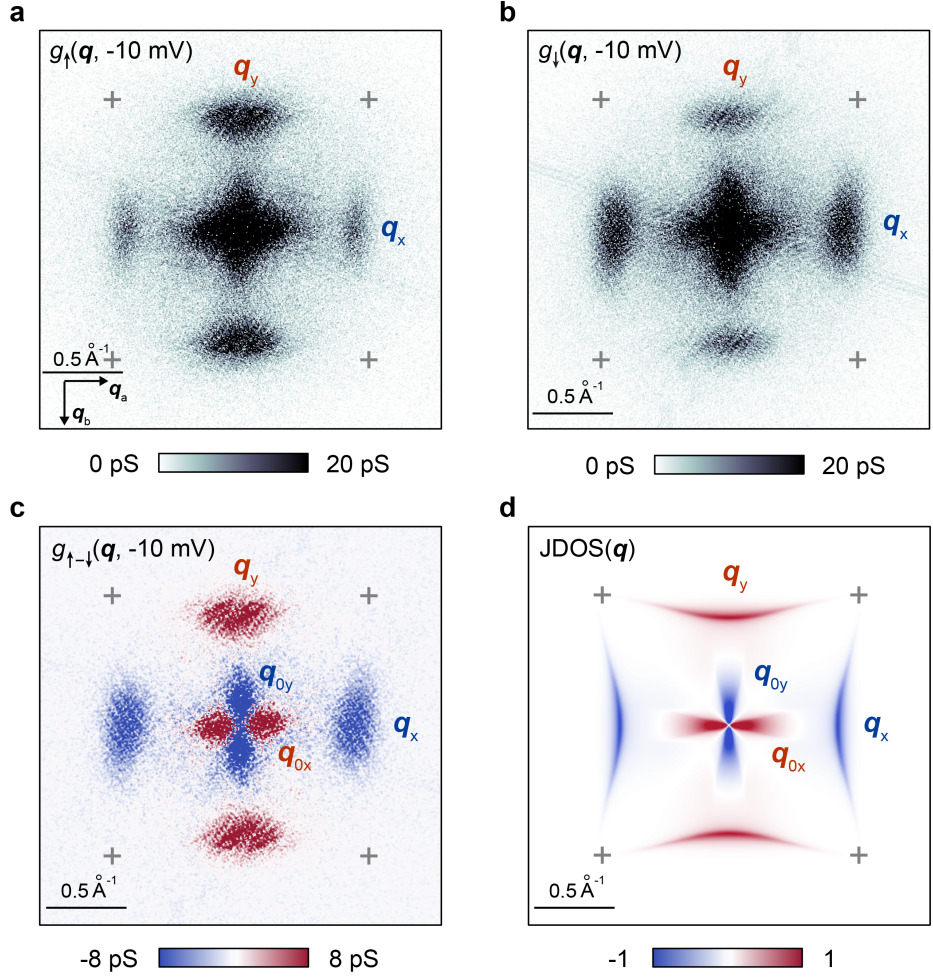

**Fig. 4 q-space imaging of spin-momentum locking. a,** $g_\uparrow(\boldsymbol{q}, V = -10\ \text{mV})$, FT of $g_\uparrow(\boldsymbol{r}, V = -10\ \text{mV})$ in Fig. 2b. The QPI pattern displays a twofold anisotropy, with dominant scattering along $\boldsymbol{q}_y$ compared to $\boldsymbol{q}_x$. Cross markers denote the surface Bragg peaks associated with the $\sqrt{2} \times \sqrt{2}$ reconstruction of Rb layer. **b,** $g_\downarrow(\boldsymbol{q}, V = -10\ \text{mV})$, FT of $g_\downarrow(\boldsymbol{r}, V = -10\ \text{mV})$ in Fig. 2d. In contrast to Fig. 4a, the QPI pattern displays a complementary anisotropy, with enhanced scattering along $\boldsymbol{q}_x$. **c,** Spin-contrast QPI pattern $g_{\uparrow-\downarrow}(\boldsymbol{q}, V = -10\ \text{mV}) = g_\uparrow(\boldsymbol{q}, V = -10\ \text{mV}) - g_\downarrow(\boldsymbol{q}, V = -10\ \text{mV})$, resolving both small-$q$ ($\boldsymbol{q}_{0x}$, $\boldsymbol{q}_{0y}$) and large-$q$ ($\boldsymbol{q}_x$, $\boldsymbol{q}_y$) scattering vectors (Extended Data Fig. 1). Red (blue) color indicates scattering intensities dominant in the spin-up (spin-down) sectors. **d,** Calculated spin-resolved QPI intensity from joint density of states (JDOS) simulation, obtained using the $k_\parallel$-filtering procedure (Extended Data Fig. 6). The striking agreement between experiment and theory identifies the spin-resolved scattering channels as a direct manifestation of spin-momentum locking in $k$ space.

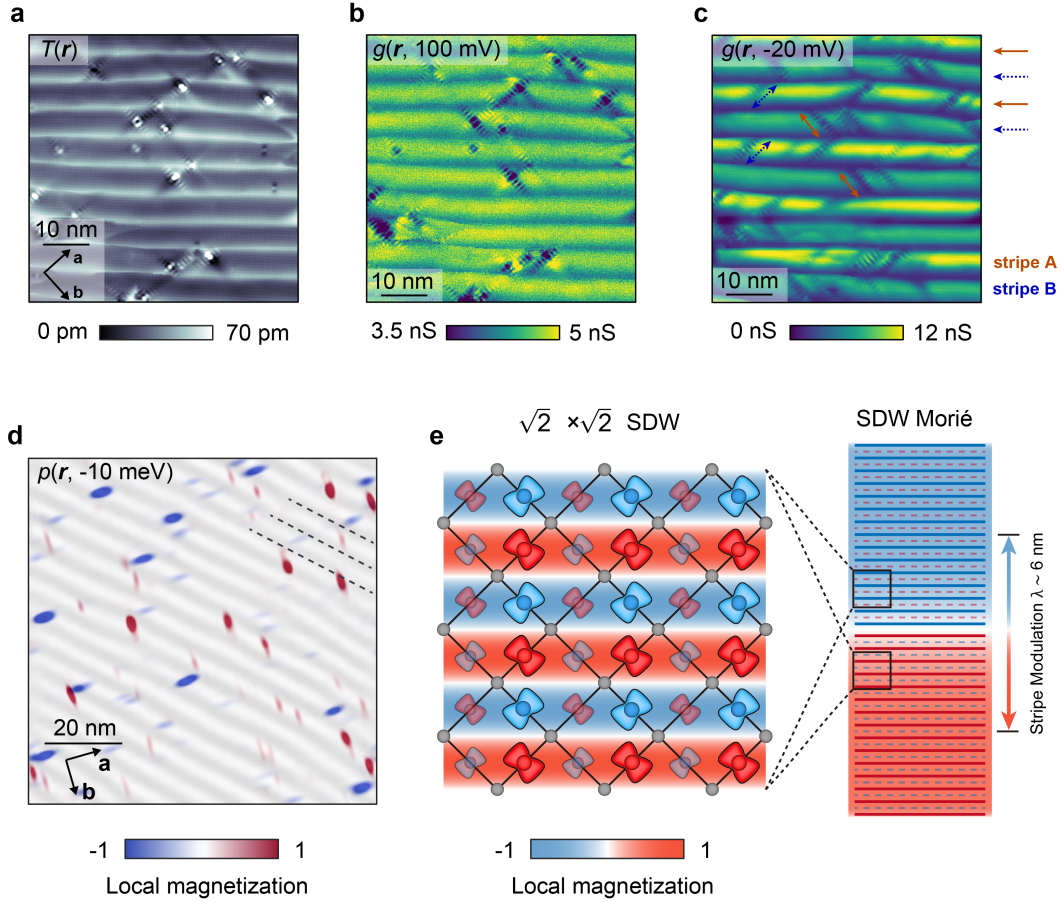

**Fig. 5 Direct visualization of unidirectional spin-stripe locking. a,** Topographic image $T(r)$, showing the stripe modulation and typical scattering centers. **b,** Measured $g(r, V = 100$ mV$)$ in the identical FOV as in Fig. 5a, exhibiting a nearly uniform unidirectional charge-stripe modulation. **c,** Measured $g(r, V = -20$ mV$)$ in the same FOV. The nearly uniform unidirectional stripe modulation in Fig. 5b evolves into alternating inequivalent A and B stripes, each associated with orthogonal scattering directions. **d,** Impurity-induced local spin polarization overlaid on the stripe modulation, alternating in sign between stripe A and stripe B. **e,** Schematic of the SDW moiré formed by the doubled periodicity (2λ) of the unidirectional charge-stripe modulation, which locks adjacent stripes into opposite local magnetizations. All measurements in this figure were performed with a setpoint of $I_s = 0.5$ nA and $V_s = 100$ mV.